\documentclass{article}
\usepackage[utf8]{inputenc}
\usepackage{graphicx}
\usepackage{amsmath}
\usepackage{mathtools}
\usepackage{authblk}

\title{Transient Dynamics of Infection Transmission in a Simulated Intensive Care Unit}

\author[1]{Christopher T. Short, PhD}
\author[1]{Matthew S. Mietchen, MPH}
\author[1]{Eric T. Lofgren, MSPH, PhD
\footnote{Corresponding Author}}
\author[\space]{for the CDC MInD-Healthcare Program}
\affil[1]{Paul G. Allen School for Global Animal Health, Washington State University, Pullman, WA}

\date{}

\begin{document}

\maketitle
\section{Abstract}
Healthcare-associated infections (HAIs) remain a public health problem. Previous work showed intensive care unit (ICU) population structure impacts methicillin-resistant \textit{Staphylococcus aureus} (MRSA) rates. Unexplored in that work was the transient dynamics of this system. We consider the dynamics of MRSA in an ICU in three different models: 1) a Ross-McDonald model with a single healthcare staff type, 2) a Ross-McDonald model with nurses and doctors considered as separate populations and 3) a meta-population model that segments patients into smaller groups seen by a single nurse. The basic reproduction number, $R_0$ is derived using the Next Generation Matrix method, while the importance of the position of patients within the meta-population model is assessed via stochastic simulation. The single-staff model had an $R_0$ of 0.337, while the other two models had $R_0$s of 0.278. The meta-population model's $R_0$ was not sensitive to the time nurses spent with their assigned patients vs. unassigned patients. This suggests previous results showing that simulated infection rates are dependent on this parameter are the result of differences in the transient dynamics between the models, rather than differing long-term equilibria.

\section{Introduction}
Healthcare-associated infections are a serious source of morbidity and mortality, and are likely to continue to be so as rates of antibiotic resistance increase. In addition to their health-related complications, these infections are also a significant burden on the resources of the healthcare system. In 2015, the Department of Health and Human Services' Hospital-Acquired Condition Reduction Program (HACRP) levied approximately \$330 million in penalties against hospitals with high infection rates \cite{Fuller}. For both reasons, reducing HAIs is a top priority for healthcare safety and quality teams. 

One such HAI, for which there has been some success in reducing rates, is methicillin-resistant \textit{Staphylococcus aureus} (MRSA). MRSA is especially difficult to treat and can be very dangerous to immune-compromised individuals and other vulnerable patients such as those in the intensive care unit (ICU) or a burn ward \cite{Kim}). MRSA is most often treated with vancomycin, a drug with a myriad of potential side effects, and a treatment failure rate of nearly 50\% \cite{Kullar}. Because of the difficulty in treating patients with MRSA once they have developed a clinical infection, a great deal of time and attention is placed on the prevention of the initial colonization of a patient with the bacteria, involving interventions such as hand hygiene or contact precautions.

In previous work \cite{Mietchen19002402}, three different methods of representing the population structure of an ICU was examined. These were 1) treating all patients as a single well-mixed group with nurses and doctors combined into a single staff type, 2) breaking nurses and doctors into two staff types with type-specific contact parameters while maintaining the well-mixed structure, and 3) representing the ICU as a meta-population, where patients are divided into groups of three with a single attending nurse per group, while the doctor sees all patients. We showed that the meta-population model had markedly lower infection rates using the same parameterization, and was generally less sensitive to changes in parameter values, suggesting models that assume a mass-action population structure might overestimate the impact of simulated interventions. 

This previous work, however, focused primarily on the long-term dynamics of these models, obtained purely by stochastic simulation. In this paper, we explore the transient dynamics of infection transmission in these systems, obtaining an analytical expression for the Basic reproduction Number ($R_0$). Additionally, we examine the impact of the initial conditions of the meta-population model in terms of where the patients are located within the ward on the first few pathogen acquisition events, a circumstance important for studies interested in the introduction of novel pathogens into the ICU, such as Ebola, MERS, or carbapenem-resistant Enterobacteriaceae.

\section{Methods}
 \begin{figure}
    \begin{center}
  \includegraphics[trim= 0cm 0cm 0cm 0cm, clip, scale=.6]{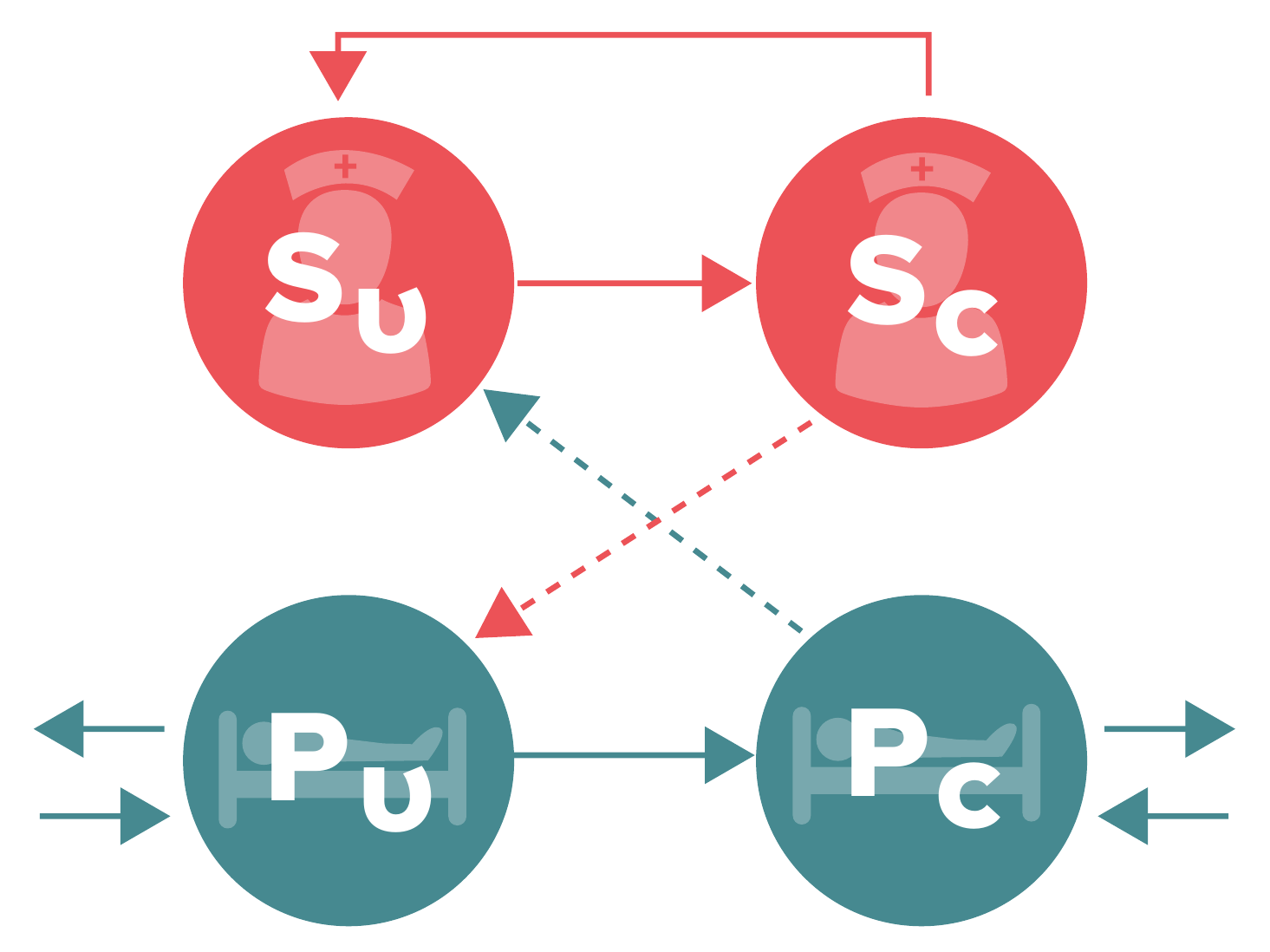}
  \caption{Schematic Representation of the Compartmental Flow of a Mathematical Model of Methicillin-resistant \textit{Staphylococcus aureus} (MRSA) Acquisition with a Single Staff Type. Solid arrows indicate possible transition states, while dashed arrows indicate potential routes of MRSA contamination or colonization. Healthcare staff are classified as uncontaminated ($S_U$) or contaminated ($S_C$), while patients are classified as uncolonized ($P_U$) or colonized ($P_C$).}
  \label{HCWSystem}
  \end{center}
\end{figure}

    \begin{figure}
    \begin{center}
  \includegraphics[scale=.6]{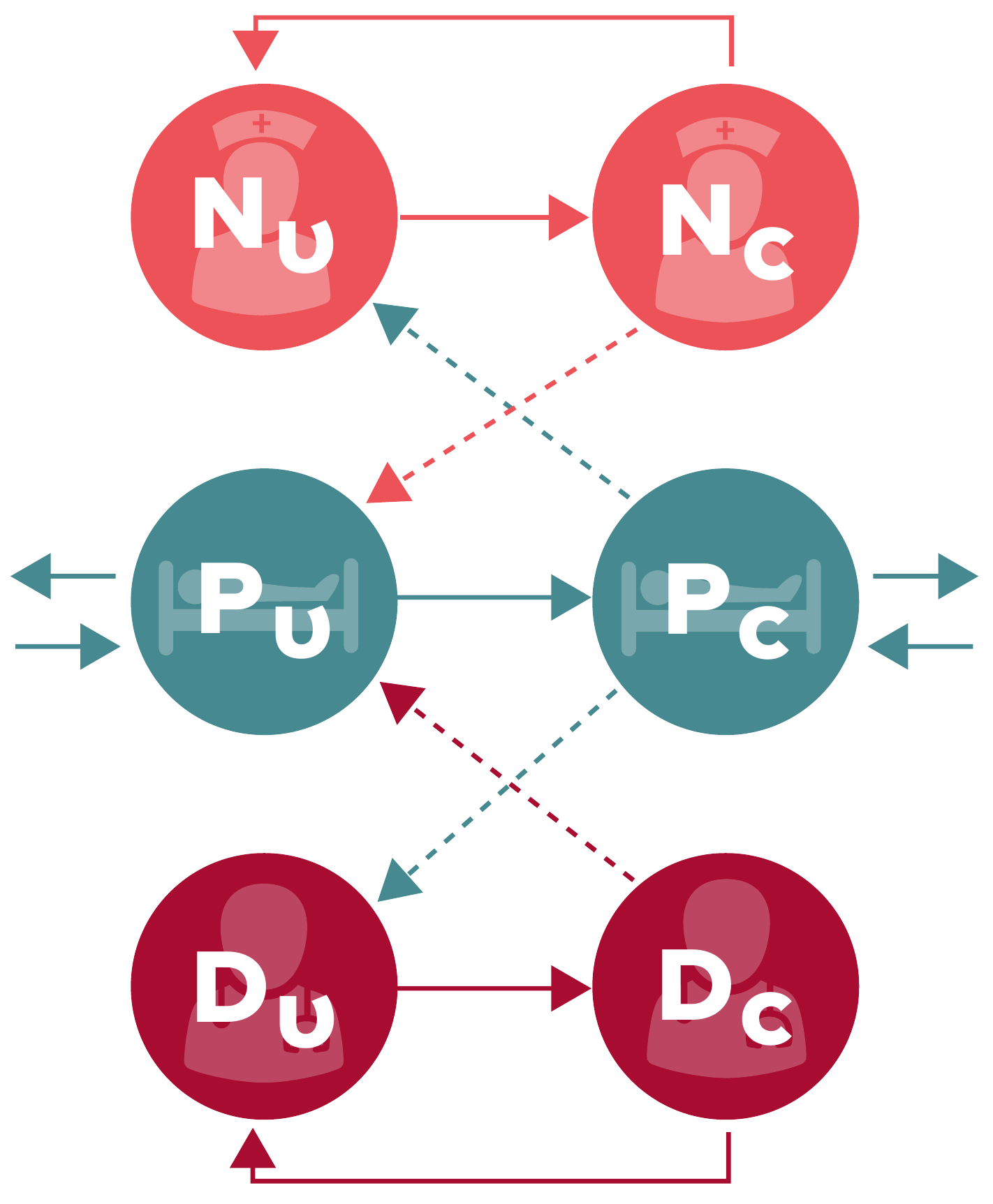}
  \caption{Schematic Representation of the Compartmental Flow of a Mathematical Model of Methicillin-resistant \textit{Staphylococcus aureus} (MRSA) Acquisition with Nurses and Intensivists Separated into Different Staff Types. Solid arrows indicate possible transition states, while dashed arrows indicate potential routes of MRSA contamination or colonization. Nurses and doctors are classified as uncontaminated ($N_U$ or $D_U$) and contaminated ($N_C$ and $D_C$), while patients are classified as uncolonized ($P_U$) or colonized ($P_C$).}
  \label{NDSystem}
  \end{center}
\end{figure}

    \begin{figure}
    \begin{center}
  \includegraphics[scale=.6]{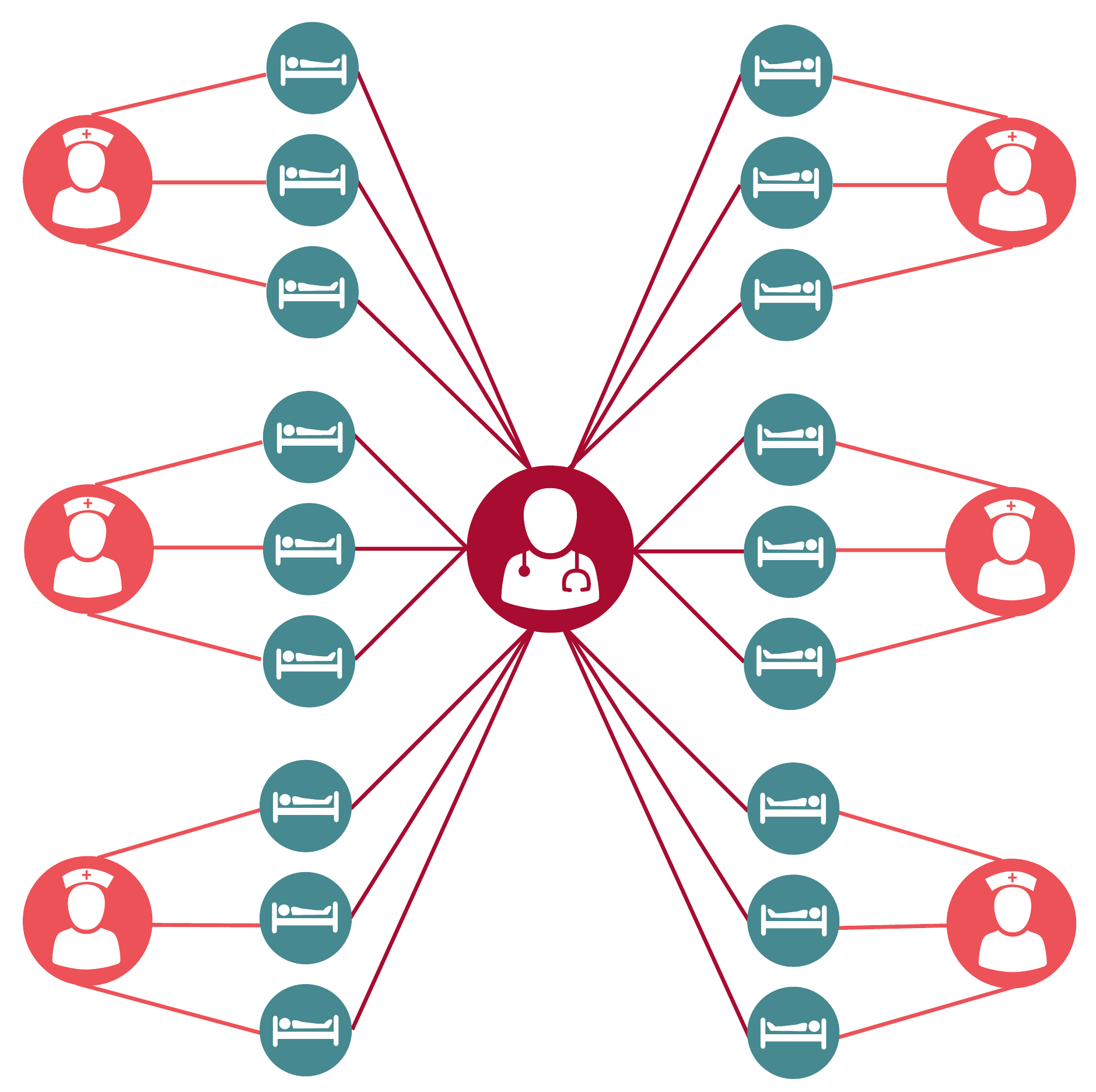}
  \caption{Schematic Representation of a Meta-population Model of Methicillin-resistant \textit{Staphylococcus aureus} (MRSA) Acquisition.  Patients (blue) are treated by a single assigned nurse (orange).  A single intensivist (red) randomly treats all patients}
  \label{Cohort}
  \end{center}
\end{figure}

\subsection{Intensive Care Unit Model}

We consider a 18-bed, single occupancy ICU, where patients are assumed, due to their critical status, to be immobile. As MRSA is not airborne, this then restricts the available MRSA transmission pathways to strictly healthcare worker (HCW) mediated patient-to-patient transmission. The role of environmental contamination is represented by modeling the contact rate in terms of "direct care tasks", which involve a healthcare worker touching either a patient or their surrounding environment, rather than in terms of patient body-contacts alone.

As previously described, several several different variations for our model ICU are analyzed.  The Single Staff Type model (Figure \ref{HCWSystem}), is the most basic model, but also the one most closely resembling the conventional Susceptible-Infected-Recovered (SIR) or Ross-McDonald models \cite{DSmith} used most commonly in the study of HAIs. Here, all 18 patients are considered one well-mixed group with 7 total health care workers, each using a weighted average of nurse- and doctor-specific contact rates. In the second model (Figure \ref{NDSystem}), the 18 patients are still viewed as a group, but the six nurses are separated from the single doctor, who now also have role-specific contact rates. Finally, we represent the ICU as a meta-population (Figure \ref{Cohort}), wherein the patients are no longer lumped together but instead placed in groups of three with a single nurse attending each group while the doctor sees all patients. An important feature of this model is the inclusion of a parameter, $\gamma$, which represents the proportion of time a nurse spends with their assigned patient group. When $\gamma = 1$ this can be considered a strict assignment, whereas the system is equivalent to a mass action model when $\gamma = \frac{1}{C}$, where C is the number of patient groups. The value of this parameter has been shown in previous simulation work \cite{Mietchen19002402} to non-linearly decrease the number of incident acquisitions of MRSA within the ICU with increasing values of $\gamma$. This formulation also reflects many of the realities of staffing, the desire for continuity of care between healthcare providers, and even the hospital built environment, where the placement of patient beds, nursing stations, etc. creates logical groupings. A table containing the parameter values can be found \ref{ParameterTable}.

For each of these models, we consider the ICU to always be at capacity as a discharge will immediately lead to an admission, maintaining a steady-state population \cite{Lofgren06192015}. Further detail on the construction, implementation and parameterization of the models may be found in \cite{Mietchen19002402}.

\subsection{Derivation of $R_0$}

The Basic Reproduction Number, $R_0$, is often of central importance when discussing the transient dynamics of a disease system, as a value less than 1 suggests that the disease is likely to die out in the long run, and pathogens with higher values of $R_0$ are, in expectation, likely to cause larger, faster outbreaks.

The Next Generation Matrix method (a worked example of which can be found in \cite{Yang2014}) separates the partial derivatives of the differential equations into two categories, one which contains only reactions involving the infected or colonized patients called F and the other containing the remaining information called V (which is composed of two pieces $V_+$ containing information regarding contamination of other individuals and $V_-$ containing decontamination information).  Subsequently the eigenvalues for the matrix $F*V^{-1}$ are calculated using the disease free equilibrium to eliminate any remaining variables. The largest (positive) eigenvalue of the matrix $F*V^{-1}$ is $R_0$ (this is somewhat similar to the Jacobian Matrix used to study the stability of an equilibrium).  The value of $R_0$ gives insight into the stability of the disease-free equilibrium.  

The use of this method involves a slight reformulation of the original model, as in the original formulation colonized patients are regularly admitted into the hospital, meaning there is no disease-free equilibrium. To correct for this, a straightforward adjustment is made where patients are still admitted, however they are never colonized on admission. This correction yields a total of three models for which we calculate $R_0$ both numerically and symbolically. An example from the single staff type model with discharges is as follows:

F=\[\begin{bmatrix}
0 & \rho \psi \frac{P_{Total}}{HW_{Total}} \\
0 & 0
\end{bmatrix}\]

V=\[\begin{bmatrix}
\mu + \theta \nu_U & 0 \\
-\rho \sigma \frac{HW_{Total}}{P_{Total}} & \iota
\end{bmatrix}\]
\subsection{Stochastic Simulation of Meta-population Initial Conditions}

The meta-population model, as it divides the patient population into strictly non-interacting groups, potentially has a starting condition not present in the other models. In any model where $\sum_{i=1}^{n} P_{C,i} \geq 2$, the placement of those patients is potentially relevant. Stochastically simulating the model using Gillespie's Direct Method \cite{Gillespie}, two initial conditions were considered - one where two patients were attended by the same nurse, and one where each patient was attended by a different nurse. These two conditions were simulated for one year assuming (as with the calculations of $R_0$ that there were no colonized admissions, and also in a more realistic circumstance where 7.79\% of admitted patients were colonized with MRSA, either from the community or elsewhere in the hospital. These simulations were performed using the StochPy package \cite{Maarleveld} and Python 3.7.

Using a panel of 1,000 runs of each model, we generated Kaplan-Meier survival curves \cite{Goel} for the time until the first MRSA acquisition and time until the third MRSA acquisition, to assess if there were any differences in the amount or timing of these early initial acquisition events. Statistical significance was assessed using a log-rank test, using the survival package \cite{Survival-Package} in R 3.5.2.

\section{Results}

\subsection{Basic Reproduction Number}

The analytical form for $R_0$ for each of the three models, as well as the specific numerical values for $R_0$ using the parameters found in the appendix are shown in Table \ref{R0Values}.


\begin{table}
\caption{Values of $R_0$ for Three ICU Models}
\begin{tabular}{l | c |r}
\label{R0Values}
Model & Numerical $R_0$ & Analytic $R_0$ \\ \hline
Single Staff Type. & 0.3367 & $\frac{\rho^2 \sigma \psi}{(\nu_u \theta + \mu) \iota}$\\ \hline
Nurse-MD Separated. & 0.2781 & $\sigma \psi \frac{\iota_d \rho_n^2 + \iota_n \rho_D^2)}{(\mu + \theta \nu_u) \iota_n \iota_d}$ \\ \hline
Meta-population & 0.2781 & $\psi \sigma \frac{6 NPT PPT \iota_n \rho_d^2 + PT \iota_d \rho_n^2}{NPT (\nu_u \theta + \mu) \iota_n PT \iota_d}$
\end{tabular}
\end{table}

\begin{figure}
    \centering
    \includegraphics[trim={9cm, 8cm, 10cm, 10cm}]{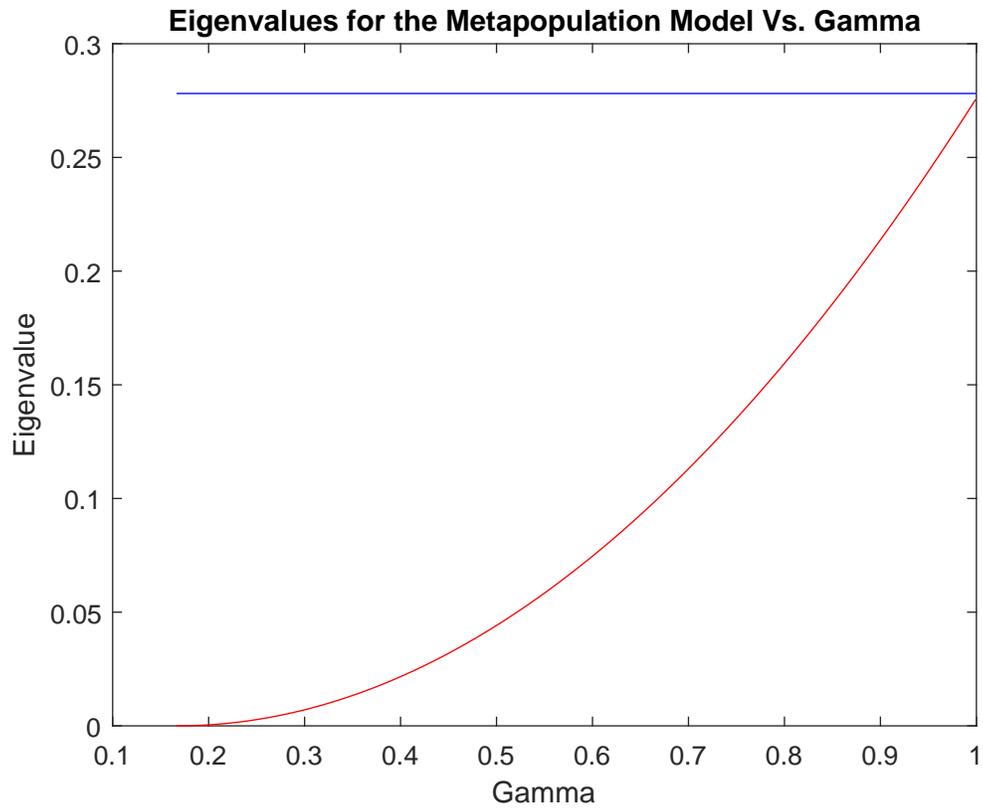}
    \caption{Values for Both Eigenvalues of the Meta-population Model for Varying Values of $\gamma$.  Note: the lines intersect at $\gamma = 1$.}
    \label{EigenvalueGraph}
\end{figure}

As with the numerical results in Mietchen \textit{et al.}, 2019, the Single Staff Type model shows a markedly higher value for $R_0$, resulting in more intensive spread, though notably, it remains below one. In contrast to previous work by Mietchen \textit{et al.}, which showed a higher infection rate in the Nurse-MD model than in the Meta-population model using stochastic simulation, these two models have the same numeric value for $R_0$, 0.2781. In the meta-population model, there are two distinct non-identical zero eigenvalues, the values of which are plotted against $\gamma$ in Figure \ref{EigenvalueGraph}. As $\gamma$ varies from $\frac{1}{6}$ (equivalent to the Nurse-MD Separated model) to $1$ (the meta-population model), it can be seen that see that: 1) at $\gamma=\frac{1}{6}$ the smaller eigenvalue becomes $0$ which also agrees with the results from the Nurse-MD separated model as there is only one non-zero eigenvalue, and 2) at $\gamma=1$ both eigenvalues intersect at $0.2781$ again leaving only one non-zero distinct eigenvalue. 

\subsection{Numerical Simulation of Meta-population Initial Conditions}

\begin{center}
    \begin{figure}
        \centering
        \includegraphics[scale=0.50]{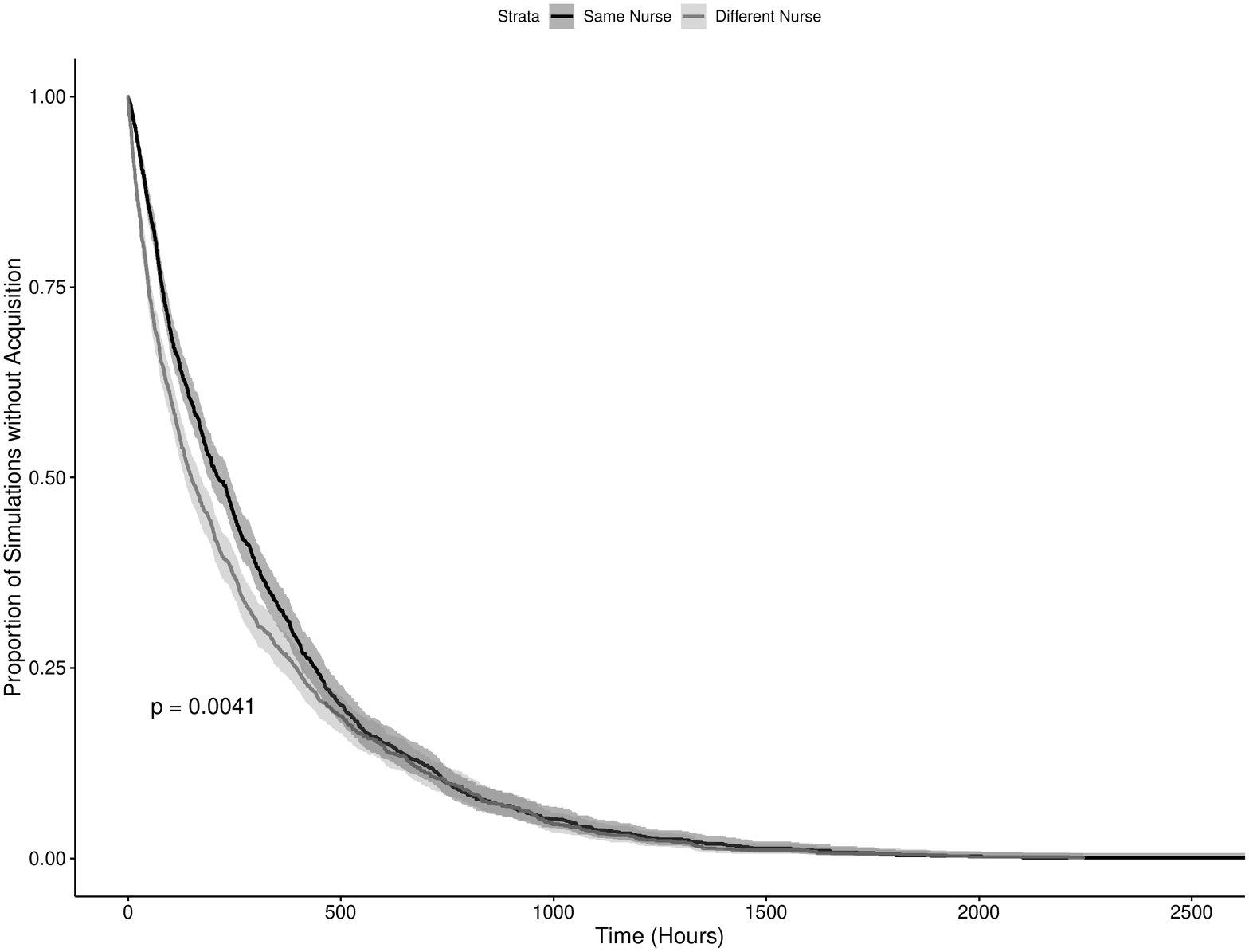}
        \caption{Time to First MRSA Acquisition in an ICU Meta-population Model with Potentially Colonized Admissions. The dark and light grey lines indicate starting conditions where two initially colonized patients are cared for by the same and different nurses respectively, with the shaded regions representing the corresponding 95\% confidence intervals.}
        \label{FigureFSFA}
    \end{figure}
\end{center}

\begin{center}
    \begin{figure}
        \centering
        \includegraphics[scale=0.50]{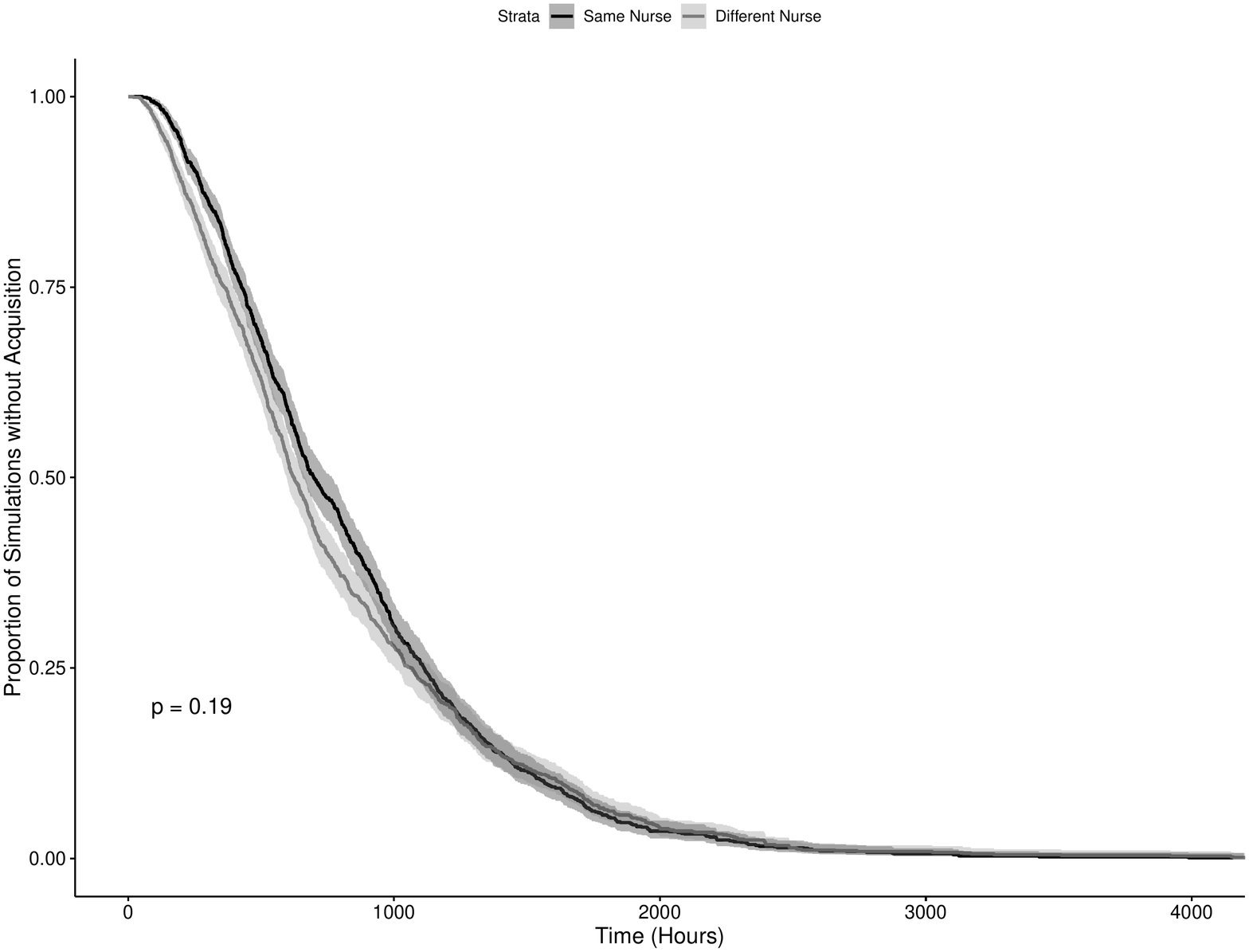}
        \caption{Time to Third MRSA Acquisition in an ICU Meta-population Model with Potentially Colonized Admissions. The dark and light grey lines indicate starting conditions where two initially colonized patients are cared for by the same and different nurses respectively, with the shaded regions representing the corresponding 95\% confidence intervals.}
        \label{FigureFSTA}
    \end{figure}
\end{center}

\begin{center}
    \begin{figure}
        \centering
        \includegraphics[scale=0.50]{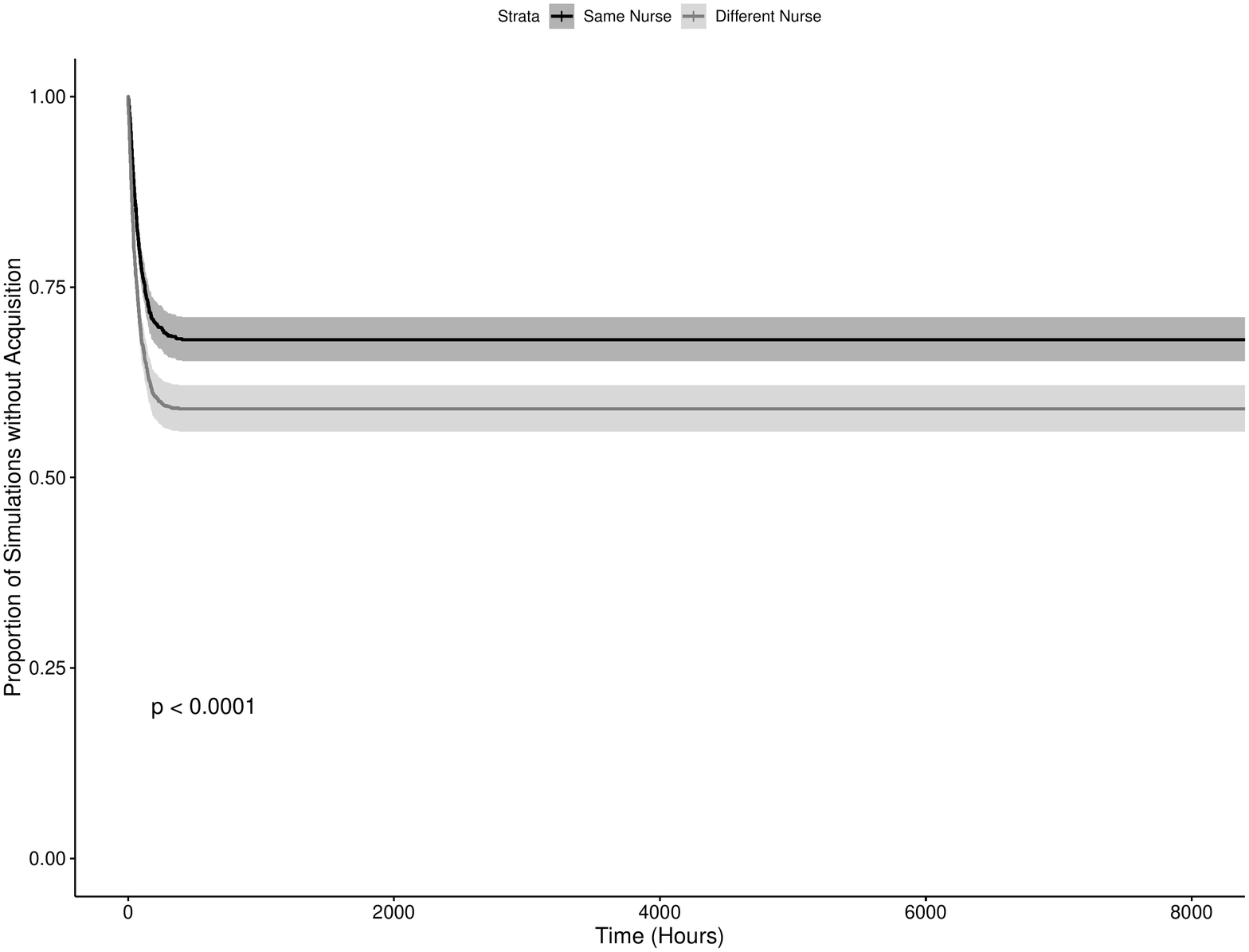}
        \caption{Time to First MRSA Acquisition in an ICU Meta-population Model with No Colonized Admissions. The dark and light grey lines indicate starting conditions where two initially colonized patients are cared for by the same and different nurses respectively, with the shaded regions representing the corresponding 95\% confidence intervals.}
        \label{FigureNCFA}
    \end{figure}
\end{center}

\begin{center}
    \begin{figure}
        \centering
        \includegraphics[scale=0.50]{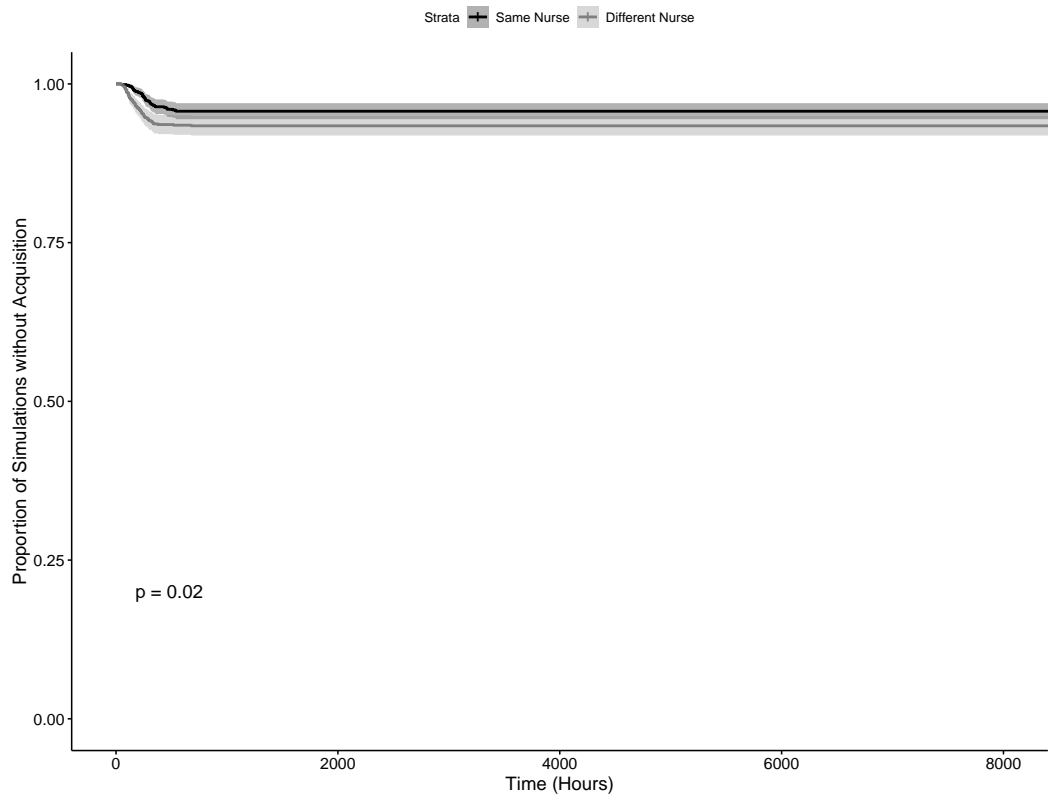}
        \caption{Time to Third MRSA Acquisition in an ICU Meta-population Model with No Colonized Admissions. The dark and light grey lines indicate starting conditions where two initially colonized patients are cared for by the same and different nurses respectively, with the shaded regions representing the corresponding 95\% confidence intervals.}
        \label{FigureNCTA}
    \end{figure}
\end{center}

The results of the numerical simulations can be seen in Figures \ref{FigureFSFA}-\ref{FigureNCTA}, showing the results for the time until the first and third acquisitions of the system with and without colonized admissions respectively. Broadly, there were statistically significant differences in the timing of the first new acquisition in both admission scenarios, with the starting condition where the two "seed" patients were cared for by the same nurse resulting in a faster new acquisition (p = 0.004 and p $>$ 0.0001 in the colonized and uncolonized admission scenarios respectively). This pattern remained significant for the third acquisition in the no colonized admissions scenario (p = 0.02), but not in the colonized admissions scenario. By the fourth acquisition, both starting conditions were statistically indistinguishable within each scenario.

Also notably, as the value of $R_0$ for these models were well below one, the majority of the simulations in the scenario with no further colonized admissions experienced rapid stochastic extinction of the pathogen. Scenarios with the two seed patients treated by different nurses were slightly less likely to go stochastically extinct, with 41.0\% of iterations (vs. 31.9\%) having a new acquisition within the hospital, and 0.66\% iterations (vs. 0.43\%) having transmission continue to a third acquisition.

\section{Conclusions and Future Work}

These results suggest that, in a model examining the introduction of a pathogen into an ICU that the transient dynamics of the system may be especially relevant. Counter to the intuition from previous work that a lower simulated infection rate in the meta-population model would correspond to a lower $R_0$, we see that this model has an equivalent $R_0$ to the Nurse-MD model. This in turn suggests that the differences seen in the models are not due to differing long-term equilibria, but rather the impact of transient dynamics and follow-on impacts from those subtle initial differences. Similarly, the difference in the initial starting placement of colonized patients within the structured population of the ICU can have considerable ramifications on the first few subsequent transmission events, though in a setting with incoming colonized patients this is quickly washed out by the noise generated by new admissions.

While at first glance the necessary modifications to obtain a disease-free equilibrium to calculate $R_0$ might seem unrealistic, this adjustment can be thought of as a representation of the transmission of a rare disease that is not yet prevalent in the community from which a hospital draws its patients. Examples of such pathogens might include emerging disease in the United States such as \textit{Candida auris} or Carbapenem-resistant Enterobacteriaceae, as well as more exotic imported diseases which have caused notable issues in healthcare settings, such as Ebola or Middle East respiratory syndrome-related coronavirus (MERS-CoV).

Also evident within the results is the suggestion that placing colonized patients under the care of the same nurse (known as "cohorting") does help control the spread of MRSA in our model.  We can also see through Figures \ref{FigureNCFA} and \ref{FigureNCTA} that barring the onslaught of incoming patients already colonized with MRSA, the idea of grouping colonized patients under the care of the same nurse is actually very effective at reducing the spread of the pathogen. The practical challenge to this practice becomes the effective detection of colonized patients, with or without evident clinical symptoms. Exploring the effectiveness of cohorting under less ideal circumstances, with imperfect diagnostics, delays in diagnostic lab results, etc. remains an area for future work. Additionally, the contrast between  figures \ref{FigureNCFA} and \ref{FigureNCTA} versus figures \ref{FigureFSFA} and \ref{FigureFSTA} suggest that this may only be true in circumstances where the admission rate of colonized patients is at or near zero. Even with a relatively low 7.79\% admission prevalence, the benefit of cohorting is quickly swamped by the colonization pressure from these new colonized patients. It is possible that a more dynamic patient admission scheme (at present incoming patients are allocated randomly) might preserve the benefits of cohorting under some circumstances. The difficulties in implementing such a scheme on a routine basis for multiple pathogens in a clinical setting are considerable, and moreso in the case of emerging pathogens. Never the less, these results point to the considerable importance of an ICU's population structure in shaping the dynamics of within-hospital infection transmission, highlighting the need for research into how these structures can be shaped by the hospital built environment, staff scheduling, hospital policy and other factors.

\section{Acknowledgements}
This work was supported by the CDC Cooperative Agreement RFA-CK-17-001-Modeling Infectious Diseases in Healthcare Program (MInD-Healthcare). 

\section{Appendix}

\subsection{Parameter Values}

\begin{tabular}{ l | c | r }
\label{ParameterTable}
Name & Value & Interpretation \\ \hline
$\rho_N$ & 3.973 & Nurse direct care tasks per hour\\ \hline
$\rho_D$ & 0.181 & Doctor direct care tasks per hour\\ \hline
$\rho$ & 4.154 & Health Care Worker direct care tasks per hour \\ \hline
$\sigma$ & 0.054 & Hand contamination probability\\ \hline
$\psi$ & 0.029 & Successful colonization of patient probability\\ \hline
$\theta$ & 0.00949 & Probability of discharge\\ \hline
$\nu_C$ & 0.0779 & Proportion of admissions colonized with MRSA\\ \hline
$\nu_U$ & (1-$\nu_C$) & Proportion of uncolonized admissions\\ \hline
$\iota_N$ & 6.404 & 11.02 nurse tasks per hour with 56.55\% compliance and 95\% efficacy\\ \hline
$\iota_D$ & 1.748 & 3.25 doctor tasks per hour with 56.55\% compliance and 95\% efficacy\\ \hline
$\iota$ & 5.74 & HCW tasks per hour with 56.55\% compliance and 95\% efficacy \\ \hline
$\tau_N$ & 2.728 & 3.30 nurse gown/glove changes per hour with 82.66\% compliance\\ \hline
$\tau_D$ & 0.744 & 0.90 doctor gown/glove changes per hour with 82.66\% compliance\\ \hline
$\tau$ & 2.445 & 2.957 gown/glove changes per hour with 82.66\% compliance \\ \hline
$\mu$ & 0.002083 & Natural decolonization rate median 20 days\\ \hline
DT & 1 & Total Number of Doctors \\ \hline
NT & 6 & Total Number of Nurses \\ \hline
PT & 18 & Total Number of Patients \\ \hline
HWT & 7 & Total Number of Health Care Workers \\ \hline
NPT & 1 & Total Number of Nurses per 'cohort' \\ \hline
PPT & 3 & Total Number of Patients per 'cohort'
\end{tabular}

\subsection{System of Equations}

\subsubsection{Health Care Worker System}
\label{HWODEs}
\begin{equation}
    \frac{dS_U}{dt}= \iota S_C +\tau S_C \frac{P_C}{P_C+P_U} - \rho \sigma S_U \frac{P_C}{P_C + P_U}
\end{equation}
\begin{equation}
    \frac{dS_C}{dt}= -\iota S_C -\tau S_C \frac{P_C}{P_C+P_U} + \rho \sigma S_U \frac{P_C}{P_C + P_U}
\end{equation}
\begin{equation}
    \frac{dP_U}{dt}=-\rho \psi P_U \frac{S_C}{S_C+S_U} + \theta \nu_U P_C +\mu P_C -\theta \nu_C P_U
\end{equation}
\begin{equation}
    \frac{dP_C}{dt}=\rho \psi P_U \frac{S_C}{S_C+S_U} - \theta \nu_U P_C -\mu P_C +\theta \nu_C P_U
\end{equation}

\subsubsection{Nurse-Doctor Model}
\label{NDODEs}
\begin{equation}
    \frac{P_U}{dt}= -\rho_N \psi P_U \frac{N_C}{N_C+N_U} - \rho_D \psi P_U \frac{D_C}{D_C+D_U} +\theta \nu_U P_C +\mu P_C - \theta \nu_C P_C
\end{equation}
\begin{equation}
    \frac{P_C}{dt}= \rho_N \psi P_U \frac{N_C}{N_C+N_U} + \rho_D \psi P_U \frac{D_C}{D_C+D_U} -\theta \nu_U P_C -\mu P_C + \theta \nu_C P_C
\end{equation}
\begin{equation}
    \frac{N_C}{dt}=-\iota_N N_C -\tau_N N_C \frac{P_C}{P_C+P_U} + \rho_N \sigma N_U \frac{P_C}{P_C+P_U}
\end{equation}
\begin{equation}
    \frac{N_U}{dt}=\iota_N N_C +\tau_N N_C \frac{P_C}{P_C+P_U} - \rho_N \sigma N_U \frac{P_C}{P_C+P_U}
\end{equation}
\begin{equation}
    \frac{dD_U}{dt}=\iota_D D_C +\tau_D D_C \frac{P_C}{P_C+P_U} -\rho_D \sigma D_U \frac{D_C}{P_C+P_U}
\end{equation}
\begin{equation}
    \frac{dD_C}{dt}=-\iota_D D_C -\tau_D D_C \frac{P_C}{P_C+P_U} +\rho_D \sigma D_U \frac{D_C}{P_C+P_U}
\end{equation}

\subsubsection{Meta-population Model}
\label{MPODEs}
\begin{equation}
\frac{dP_{Ui}}{dt} = -\rho_N \psi  \gamma \frac{N_{Ci} P_{Ui}}{N_{Ci} + N_{Ui}} -\rho_N \psi  \frac{1-\gamma}{5} \frac{N_{Cj} P_{Ui}}{N_{Cj} + N_{Uj}} -\rho_D \psi \frac{D_C P_{Ui}}{D_C + D_U}
\nonumber
\end{equation}
\begin{equation}
-\theta \nu_C P_{Ui} + \theta \nu_U P_{Ci} + \mu P_{Ci} - \theta P_{Ui} \nu_C
\end{equation}

\begin{equation}
\frac{dP_{Ci}}{dt} = \rho_N \psi  \gamma \frac{N_{Ci} P_{Ui}}{N_{Ci} + N_{Ui}} +\rho_N \psi  \frac{1-\gamma}{5} \frac{N_{Cj} P_{Ui}}{N_{Cj} + N_{Uj}} +\rho_D \psi \frac{D_C P_{Ui}}{D_C + D_U}
 \nonumber
 \end{equation}
 \begin{equation}
 +\theta \nu_C P_{Ui} - \theta \nu_U P_{Ci} - \mu P_{Ci} -\theta P_{Ci} \nu_C
 \end{equation}

\begin{equation}
\frac{dN_{Ui}}{dt} = \iota_N N_{Ci} + \tau_N \gamma \frac{N_{Ci} P_{Ci}}{P_{Ui} + P_{Ci}} + \tau_N \frac{1-\gamma}{5} \frac{N_{Ci} P_{Cj}}{P_{Uj} +P_{Cj}} - \nonumber
\end{equation}
\begin{equation}
\rho_N \sigma \gamma \frac{N_{Ui} P_{Ci}}{P_{Ui} + P_{Ci}} - \rho_N \sigma \frac{1-\gamma}{5}\frac{N_{Ui} P_{Cj}}{P_{Uj} + P_{Cj}}
\end{equation}

\begin{equation}
\frac{dN_{Ci}}{dt} = -\iota_N N_{Ci} - \tau_N \gamma \frac{N_{Ci} P_{Ci}}{P_{Ui} + P_{Ci}} - \tau_N \frac{1-\gamma}{5} \frac{N_{Ci} P_{Cj}}{P_{Uj} +P_{Cj}} + \nonumber
\end{equation}
\begin{equation}
\rho_N \sigma \gamma \frac{N_{Ui} P_{Ci}}{P_{Ui} + P_{Ci}} + \rho_N \sigma \frac{1-\gamma}{5}\frac{N_{Ui} P_{Cj}}{P_{Uj} + P_{Cj}}
\end{equation}

\begin{equation}
\frac{dD_U}{dt} = \iota_D D_C + \tau_D D_C \sum_{i=1}^6 \frac{P_{Ci}}{P_{Ui} + P_{Ci}} - \rho_D \sigma D_U \sum_{i=1}^6 \frac{P_{Ci}}{P_{Ui} + P_{Ci}}
\end{equation}

Note: $\gamma$ represents a percentage of time a nurse would spend tending patients not assigned to him/her.  If $\gamma = \frac{1}{6}$ the system reduces to the Nurse-MD model, and $\gamma = 1$ is the full meta-population model where no nurse tends to unassigned patients.

\bibliographystyle{unsrt}
\bibliography{Bibliography.bib}

\end{document}